\documentclass[layout=twocolumn,journal=aamick, manuscript=article]{achemso}

\usepackage{blindtext}
\usepackage{dcolumn}
\usepackage{ulem}
\usepackage[utf8]{inputenc}
\usepackage{amsmath,amssymb}
\usepackage{amsfonts} 
\usepackage{graphicx}
\usepackage{xcolor}
\usepackage{hyperref}
\usepackage{mathtools, cuted}
\usepackage{lipsum,color}

\newcommand{\vect}[1]{\boldsymbol{#1}}
\newcommand*{\citen}[1]{%
(\begingroup
    \romannumeral-`\x 
    \setcitestyle{numbers}%
    \cite{#1}%
  \endgroup)}

\title{ Enhanced Photo-Excitation and Angular-Momentum Imprint of Gray Excitons in WSe$_{2}$ Monolayers by Spin-Orbit-Coupled Vector Vortex Beams}

\author{Oscar Javier Gomez Sanchez$^{\#}$}
\affiliation{Department of Electrophysics, National Yang Ming Chiao Tung University, Hsinchu 300, Taiwan}

\author{Guan-Hao Peng$^{\#}$}
\email{bnm852i@gmail.com}
\affiliation{Department of Electrophysics, National Yang Ming Chiao Tung University, Hsinchu 300, Taiwan}

\author{Wei-Hua Li}
\affiliation{Department of Electrophysics, National Yang Ming Chiao Tung University, Hsinchu 300, Taiwan}

\author{Ching-Hung Shih}
\affiliation{Institute of Electronics, National Yang Ming Chiao Tung University, Hsinchu 300, Taiwan}

\author{Chao-Hsin Chien}
\affiliation{Institute of Electronics, National Yang Ming Chiao Tung University, Hsinchu 300, Taiwan}

\author{Shun-Jen Cheng}
\email{sjcheng@nycu.edu.tw }
\affiliation{Department of Electrophysics, National Yang Ming Chiao Tung University, Hsinchu 300, Taiwan}

\keywords{gray exciton, two-dimensional materials, transition-metal dichalcogenide, twisted light, vector vortex beam, WSe$_2$.}
\let\oldmaketitle\maketitle
\let\maketitle\relax

\begin{document}

\twocolumn[
\begin{@twocolumnfalse}

\oldmaketitle

\noindent $^{\#}$ These authors contributed equally to this work and their names are
listed by alphabetical order.

\begin{abstract}
A light beam can be spatially structured in the complex amplitude to possess orbital angular momentum (OAM), which introduces an extra degree of freedom alongside the intrinsic spin angular momentum (SAM) associated with circular polarization. Furthermore, super-imposing two such twisted light (TL) beams with distinct SAM and OAM produces a vector vortex beam (VVB) in non-separable states where not only complex amplitude but also polarization are spatially structured and entangled with each other.  In addition to the non-separability, the SAM and OAM in a VVB are intrinsically coupled by the optical spin-orbit interaction and constitute the profound spin-orbit physics in photonics.   
In this work, we present a comprehensive theoretical investigation, implemented on  the first-principles base, of the intriguing light-matter interaction between VVBs and WSe$_{2}$ monolayers (WSe$_{2}$-MLs), one of the best-known and promising two-dimensional  (2D)  materials  in optoelectronics dictated by excitons, encompassing bright exciton (BX) as well as various dark excitons (DXs). 
One of the key findings of our study is that a substantial enhancement of the photo-excitation of gray excitons (GXs), a type of spin-forbidden dark exciton, in a WSe$_2$-ML can be achieved through the utilization of a 3D-structured TL with the optical spin-orbit interaction. Moreover, we show that a spin-orbit-coupled VVB surprisingly allows for the imprinting of the carried optical information onto gray excitons in 2D materials, which is robust against the decoherence mechanisms in materials.  This suggests a promising method for deciphering the transferred angular momentum from structured lights to excitons.
\end{abstract}
\end{@twocolumnfalse}
]

A spatially structured light beam with a cylindrically twisted phase front introduces quantized orbital angular momenta (OAM), $L _{z} = \ell \hbar$, which serves as an extra  degree of freedom for light alongside spin angular momentum (SAM), $S _{z} = \sigma \hbar$, where $\sigma = \pm 1$ represents the right- or left-handed circular polarization of light. \cite{COULLET1989403,allen1992orbital,PhysRevLett.80.3217,PhysRevLett.88.053601}
Such a cylindrically structured beam, also referred to as twisted light (TL) or optical vortex (OV), is denoted by $\vert \sigma , \ell \rangle$ and characterized by an unbounded quantum number $\ell$ has been demonstrated advantageous in a variety of advanced photonic and quantum applications, ranging from optical tweezers, \cite{padgett2011tweezers,gong2018optical} optical trapping, \cite{otte2020optical,yang2021optical}  high-resolution optical microscope, \cite{tan2010high,zhang2016perfect,ritsch2017orbital} optical communication, \cite{wang2012terabit,xie2018ultra} to high dimensional quantum information. \cite{mair2001entanglement,fickler2012quantum,boyd2016quantum,erhard2018twisted} 
Besides, the co-existence of SAM and OAM in a structured light beam gives rise to intriguing optical spin-orbit-coupled phenomena, \cite{allen1996spin,eismann2019spin,bliokh2015spin} including photonic spin Hall effect, \cite{bliokh2015quantum,ling2017recent,bahari2021photonic,kim2023spin} spin-based plasmonics, \cite{yu2021spin} photonic wheel, \cite{banzer2013photonic,aiello2015transverse} optical transverse spin, \cite{shao2018spin} and longitudinal field of light. \cite{zhao2007spin,bliokh2015transverse,shi2018structured,forbes2021orbital,forbes2021relevance}

Furthermore, a structured light beam can be tailored by the controlled superposition of TLs with distinct SAM and OAM, expressed by
\begin{align}
\resizebox{\linewidth}{!}{$ 
|\sigma , \ell, \sigma' , \ell' ; \alpha,\beta \rangle = \cos \left( \beta/2 \right) |\sigma , \ell \rangle  + e^{i\alpha} \sin \left( \beta/2 \right) | \sigma' , \ell'  \rangle \, ,$}
\label{superposition}
\end{align}
where $\alpha$ ($\beta$) determines the relative phase (weight) of the two TL components in the superposition state and represents the azimuthal (polar) angle in the Poincaré sphere representation. \cite{milione2011higher,naidoo2016controlled,liu2017generation,Huang17} The structured light described by Eq. (\ref{superposition}) forms a vector vortex beam (VVB) in non-separable states, where not only the complex amplitude but also the polarization of light are spatially structured and entangled with each other. \cite{mclaren2015measuring,naidoo2016controlled,Huang17, Forbes2021structured,lu2022generation} 
The exceptional characteristics of VVBs as light sources have been demonstrated to enable advanced photonics applications, \cite{zhang2016vectorial} particle acceleration, \cite{wong2010direct,chen2015structured} vector beam multiplexing communication, \cite{milione20154,fang2018spin} high dimensional quantum entanglement, \cite{tang2016entanglement,krenn2017orbital} and vector vortex quantum steering. \cite{slussarenko2022quantum} 
The non-separability of the SAM and OAM, further coupled by the optical spin-orbit interaction (SOI), in a VVB embodies the profound spin-orbit physics of optics and naturally affect its interaction with matters, which, however, remain largely unexplored so far.   Following the rapid advancement in the TL-based optics, it is certainly crucial to investigate the physics of the interaction between structured lights and seek the emergent nano-materials suited for the prospective TL-based optoelectronics. \cite{rosen2022interplay,simbulan2021selective,ji2020photocurrent,session2023optical,feng2022twisted} 
Specifically, identifying materials that facilitate efficient transfer of optical angular momenta from structured light and enable deciphering of transmitted optical information is highly desirable for prospective TL-based optoelectronics.

Atomically thin transition-metal dichalcogenide monolayer (TMD-ML) is one of the most promising optoelectronic 2D materials with superior light-matter interactions that are dictated by excitons. \cite{mak2010atomically,bernardi2013extraordinary,chernikov2014exciton,he2014tightly,manzeli20172d}  In TMD-MLs, excitons are strongly bound by the enhanced Coulomb interaction, leading to the  atypical band dispersion and exciton fine structures associated with the diverse degrees of freedom inherent in excitons, including spin and valley properties as well as the center-of-mass motion of exciton. \cite{yu2014valley,sauer2021optical,PhysRevResearch.3.043198,ma2022coherent,li2023key} 
The  pronounced     exciton fine structure of a TMD-ML enables the unambiguous spectral resolution of diverse exciton complexes, such as the bright exciton and various dark exciton states, \cite{chen2018coulomb} each possessing distinct degrees of freedom.  In darkish W-based TMD-MLs, e.g. WSe$_2$, \cite{molas2017brightening} the intravalley repulsive exchange energy combined with the conduction band splitting shifts the dipole-allowed bright exciton states upwards by tens of meV and leave the spin-forbidden dark exciton doublet as the excitonic ground states. \cite{zinkiewicz2020neutral} Furthermore, the lowest doublet of dark excitons undergoes valley-mixing, resulting from weak intervalley exchange interaction, and exhibits a slight energy splitting, yielding a completely dark exciton and a slightly optically active state known as a gray exciton (GX). \cite{PhysRevLett.119.047401,robert2017fine,robert2020measurement}

Notably, GXs have recently garnered significant attention due to their  possession of the both advantages from bright excitons (BXs) as well as dark excitons (DXs), i.e. long lifetime and brightness. \cite{schneider2020direct,ma2022coherent}
These characteristics are highly desirable for future exciton-based quantum technologies and devices. \cite{feierabend2017proposal,jiang2018microsecond} Nevertheless, optically accessing the GX states remains a non-trivial task and usually needs the additional aid of external fields or post-processed structures of samples, such as in-plane magnetic fields, \cite{zhang2017magnetic,molas2017brightening,robert2020measurement} plasmonic fields, \cite{zhou2017probing} or photonic crystals in close proximity. \cite{tang2019long,ma2022coherent} 
Despite the out-of-plane dipole and the expected light emission along the plane of 2D materials, direct observation of GXs in TMD-MLs has been shown to be achievable by using high numerical aperture objectives in both regular photoluminescence (PL) spectroscopies \cite{luo2020exciton,li2019direct} where the detectors are set in the normal direction to the 2D materials, and angle-resolved optical spectroscopies.
The fascinating attributes of TL have recently stimulated a few pioneering investigations concerning their interactions with BXs in 2D systems. \cite{shigematsu2013orbital,PhysRevB.93.045205,yang2021steering,simbulan2021selective,pattanayak2022steady,PhysRevB.105.205202,kesarwani2022control,peng2022tailoring,zhang2023single,pattanayak2022probing} However, beyond scalar optical vortex beams or TLs, the interplay between VVBs and excitons in 2D materials remains an appealing but yet largely unexplored area.

In this study, we present a comprehensive theoretical investigation based on first-principles, focusing on the interaction between spin-orbit-coupled VVBs and exciton states in a WSe$_2$-ML, including both BXs and GXs.
 We reveal that structured lights can serve as an exceptional light source enabling optically enhance the photo-excitation of GXs in a WSe$_{2}$-ML through the coupling of the longitudinal field component associated with the SOI.  Furthermore, we show that a spin-orbit-coupled VVB enables the imprinting of optical information onto the optical transitions of GXs in the 2D materials.

\section*{RESULTS AND DISCUSSION }

\begin{figure*}
	    \centering
        \includegraphics[width=\textwidth]{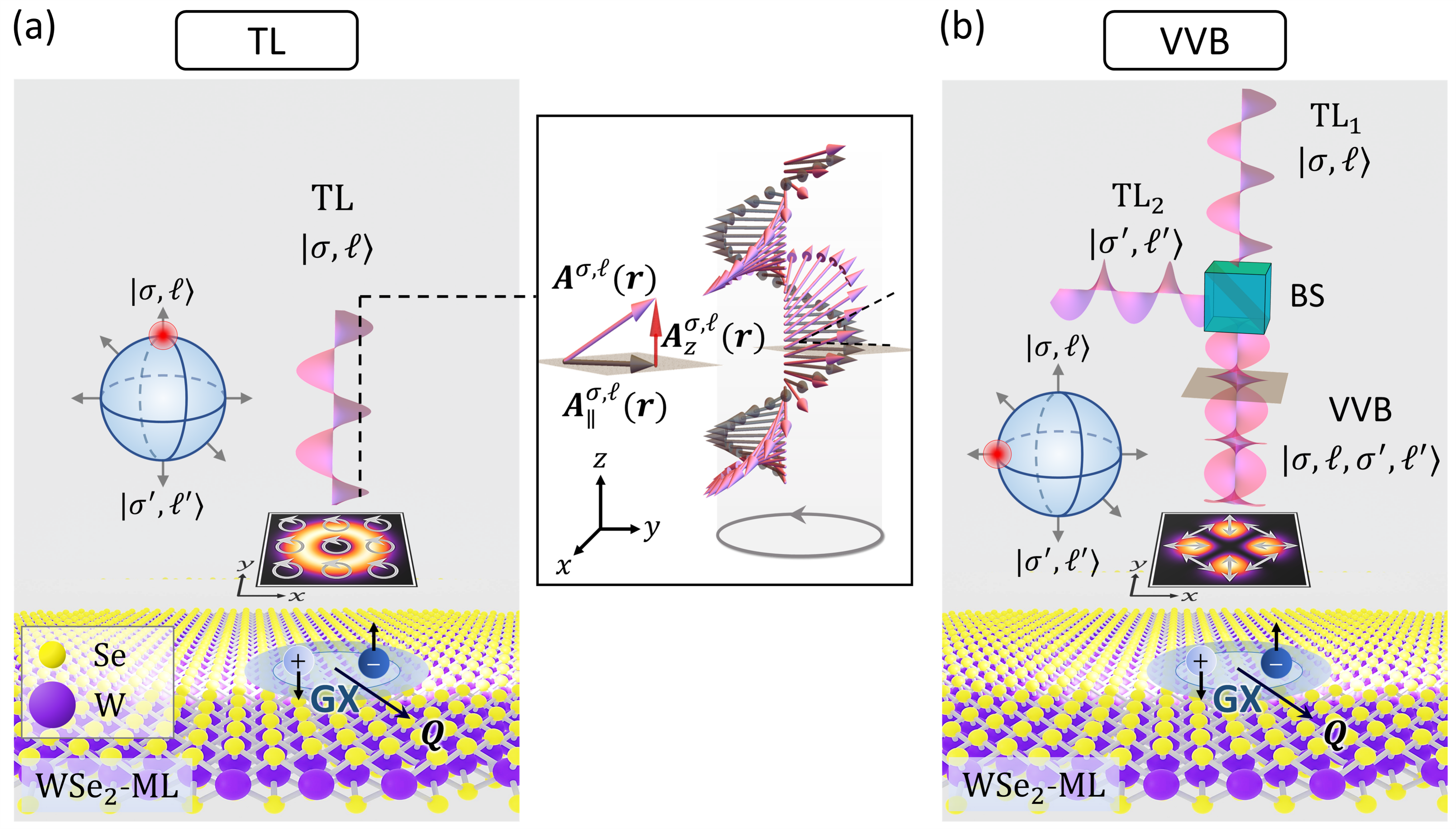}
	    \caption{(a) The schematic of a single twisted light (TL) in the state $\vert \sigma, \ell \rangle$ normally incident on a WSe$_{2}$ monolayer (WSe$_{2}$-ML), where $\sigma$ and $\ell$ represent the spin angular momentum (SAM) and orbital angular momentum (OAM) carried by the TL, respectively. The panel on the right-hand side shows that the polarization of a TL (pink arrow line) is not purely transverse (gray arrow line), $\vect{A}_{\parallel}^{\sigma, \ell} (\vect{r}) = e^{iq_0z}{A}_{\parallel}^{\ell} (\vect{\rho}) \hat{\vect{\varepsilon}}_{\parallel}^{\sigma}$, but also contains a small longitudinal field component (red arrow line), $\vect{A}_z^{\sigma, \ell} (\vect{r}) = e^{iq_0z}{A}_{z}^{\sigma, \ell} (\vect{\rho}) \hat{\vect{\varepsilon}} _{z}$, resulting from the optical spin-orbit interaction (SOI). Below the TL, the small square panel displays the contour plot of the squared magnitude of the longitudinal component of the vector potential and the in-plane circular polarizations (white arrow lines) of the TL, where the polarization remains fixed over the in-plane. The schematic on the left-hand side illustrates that the TL state, $\vert \sigma, \ell \rangle$, is located at the north pole of the Poincar\'e sphere. (b) The schematic of a vector vortex beam (VVB) superimposed by two TLs, $\vert \sigma, \ell \rangle$  and $\vert \sigma', \ell' \rangle$, normally incident on a WSe$_2$-ML. BS stands for beam splitter. Below the VVB, the small square panel displays the contour plot of the squared magnitude of the longitudinal component of the vector potential and the in-plane polarizations of the VVB, both of which are spatially varied, unlike a TL. The VVB here is a state located at the equator of the Poincar\'e sphere, i.e. $\vert\sigma,\ell,\sigma',\ell'\rangle\equiv\vert\sigma,\ell,\sigma',\ell';\alpha = 0,\beta=\pi/2\rangle $ (see Eq. (\ref{superposition}) for the definition).}
	    \label{Fig1}
\end{figure*}

The vector potential of a circularly polarized Laguerre-Gaussian (LG) TL in the   fundamental radial mode $p=0$ under the Coulomb gauge is 
$\vect{A} ^{\sigma, \ell} (\vect{r}) = e ^{i q _{0} z}\vect{A}^{\sigma, \ell}(\vect{\rho}) = e ^{i q _{0} z} [\hat{\vect{\varepsilon}} _{\parallel}^{\sigma} A_{\parallel}^{\ell}(\vect{\rho}) + \hat{\vect{\varepsilon}} _{z}  A_{z}^{\sigma, \ell} (\vect{\rho}) ]$, (see Supporting Information for details) being a 3D-structured light with the both transverse and longitudinal components \cite{quinteiro2017twisted}  as illustrated in Fig. \ref{Fig1}(a), where $\vect{r}=(x,y,z)=(\vect{\rho}, z)$ is the  position vector, $\hat{\vect{\varepsilon}} _{\parallel} ^{\sigma} = \frac{1}{\sqrt{2}}(\hat{\vect{x}} + i \sigma  \hat{\vect{y}})$ is the transverse polarization vector labelled by the optical helicity $\sigma=\pm 1$, $\hat{\vect{\varepsilon}} _{z} = \hat{\vect{z}}$ is the longitudinal polarization vector, $\ell=0,\pm 1, \pm2, \pm3,...$ is the index of the azimuthal mode of light, \cite{romero2002quantum}  $q _{0} = 2 \pi n _{rf} / \lambda$ is the wave number of light propagating along the $z$-direction, $\lambda$ is the wavelength of light in vacuum, and $n _{rf}$ is the refractive index of material. 
In the formalism of vector potential, the spin part of light is associated with the transverse vector of polarization, $\hat{\vect{\varepsilon}} _{\parallel}^{\sigma}$, and the orbital part of light is characterized by the scalar function of the vector potential, $A _{\parallel} ^{\ell} (\vect{\rho})$.
Throughout this work, 
we consider the TL with the beam waist, $w_0=1.5 \ \mu$m. The vector potential of a structured light beam in general varies along, besides the in-plane direction, the $z$-direction as well. However, within a narrow $z$-range characterized by the Rayleigh length, $z_R$, the $z$-dependence of vector potential is very weak and one can adopt the long Rayleigh length approximation to neglect the $z$-dependence of vector potential. Without loss of generality, we simply consider $n_{rf}=1$ for most studies in this work. For the TL with $w _{0} = 1.5 \, \mu$m and the resonant frequency to the exciton transition energy of $E ^{X} _{B \pm , \vect{0}} = 1.7$ eV, \cite{jones2013optical,huang2016probing} the Rayleigh length for the TL is estimated as $z_R=11.81\,\mu$m according to the relationship between light speed $c$, light wavevector $q_0$, resonant excitation energy $\hbar c q_0 = E ^{X} _{B \pm, \vect{0}}$, and the definition of Rayleigh length $z_R= q_0 w_0^2/2$. \cite{jones2013optical,huang2016probing} 
Considering the nm-scaled thickness of 2D materials $d << z_R$ \cite{huang2015large}, we adopt the long Rayleigh length approximation and disable the variable $z$ in the vector potential throughout this work. As stated in Sec. SII.B of the Supporting Information, the TL with a smaller Rayleigh length, $z _{R}$, will give rise to the enhanced longitudinal component and higher transition rates for GXs.

For the integration with the light-matter interaction based on the exciton band structures of 2D materials,  it is necessary to transform the vector potentials into the angular spectrum representation through a 2D Fourier transform. 
The Fourier transform of the complex transverse component of 
$\vect{\mathcal{A}}^{\sigma, \ell} (\vect{q}_{\parallel}) \equiv \frac{1}{\Omega}  \int d ^{2} \vect{\rho} \, \vect{A}^{\sigma, \ell}(\vect{\rho}) e^{-i\vect{q}_{\parallel} \cdot \vect{\rho}} = \hat{\vect{\varepsilon}} _{\parallel}^{\sigma} \mathcal{A}_{\parallel}^{\ell}(\vect{q}_{\parallel}) + \hat{\vect{\varepsilon}} _{z} \mathcal{A}_{z}^{\sigma, \ell} (\vect{q}_{\parallel})$
is derived as 
\begin{align}
{\mathcal{A}}_{\parallel}^{\ell}(\vect{q}_{\parallel}) = \Tilde{F}_{|\ell|}(q_\parallel)e^{i\ell\phi_{\vect{q}}}\, , \label{Aqp}
\end{align} 
  and the longitudinal one as
\begin{align}
\mathcal{A}_{z}^{\sigma, \ell}(\vect{q}_{\parallel}) \approx - (\hat{\vect{\varepsilon}}_{\parallel}^{\sigma} \cdot \hat{\vect{q}}) \mathcal{A} _{\parallel} ^{\ell} (\vect{q} _{\parallel}) \label{Aqz}\, ,
\end{align} 
as detailed in section SII of Supporting Information, where $\Omega$ is the area of the system, $\hat{\vect{q}}=\vect{q}/|\vect{q}|$ with $\vect{q}=(\vect{q}_\parallel,q_0)$, $\vect{q}_\parallel=(q_x,q_y)$, and $\phi_{\vect{q}} = \tan ^{-1} (q_y/q_x)$. 
The expression for the complex-valued radial function can be represented as $\tilde{F} _{|\ell|} (q _{\parallel}) = (-i) ^{|\ell|} F _{|\ell|} (q _{\parallel})$, where the detailed description of the amplitude function $F _{|\ell|} (q _{\parallel})$ is provided in Supporting Information. \cite{peng2022tailoring}
  In turn, the vector potential as a function of coordinate position in the real space can be expressed as $\vect{A}^{\sigma, \ell}(\vect{\rho})=\sum_{\vect{q}_\parallel} \vect{\mathcal{A}}^{\sigma, \ell} (\vect{q}_{\parallel}) e^{i\vect{q}_{\parallel} \cdot \vect{\rho} }$, via the inverse Fourier transform. \cite{simbulan2021selective,peng2022tailoring}
  
 The appearance of $\mathcal{A} _{\parallel} ^{\ell} (\vect{q}_{\parallel})$ in Eq. (\ref{Aqz}) accounts for that the longitudinal field in a TL fully inherits the OAM-encoded transverse spatial structures described by Eq. (\ref{Aqp}).
 Notably, the term $(\hat{\vect{\varepsilon}}_{\parallel}^{\sigma} \cdot \hat{\vect{{q}}})  =(\hat{\vect{\varepsilon}}_{\parallel}^{\sigma} \cdot \vect{q}_{\parallel})/q $ appearing in Eq. (\ref{Aqz})  manifests itself as the optical SOI that couples  the  optical spin ($\hat{\vect{\varepsilon}}_{\parallel}^{\sigma}$) and the  in-plane momentum component (${\vect{{q}}_{\parallel}}$) carried by the longitudinal field. Alternatively,  $(\hat{\vect{\varepsilon}}_{\parallel}^{\sigma} \cdot \hat{\vect{{q}}}) =\frac{\sin \theta_{\vect{q}}}{\sqrt{2} } e^{i \sigma \phi_{\vect{q}}} $ with $\sin \theta_{\vect{q}}\equiv \frac{q_{\parallel}}{\sqrt{q_{\parallel}^2 +q_0^2} }$ can be expressed in the spherical  coordinates,  showing that the optical spin $\sigma$ is fully transferred to the longitudinal field and the strength of optical SOI increases with increasing $ q_{\parallel} $. 
Combining $\mathcal{A}_{\parallel}^{\ell}(\vect{q}_{\parallel})$ and $(\hat{\vect{\varepsilon}}_{\parallel}^{\sigma} \cdot \hat{\vect{{q}}})$, the longitudinal field expressed by Eq. (\ref{Aqz})  is shown imprinted by $\hbar (\sigma+\ell) \equiv \hbar J $, which is the total angular momentum (TAM) of TL in the paraxial regime. \cite{allen1996spin}

In Supporting Information, Figure S1(a) and (b) [(c) and (d)] show the squared magnitude, the real part, and the imaginary part of the complex vector  potential  $\mathcal{A}_{\parallel}^{\ell} (\vect{q}_{\parallel})$ [$\mathcal{A}_{z}^{\sigma, \ell} ( \vect{q}_{\parallel} )$], as functions of $\vect{q} _{\parallel}$ for the polarized TLs in the LG modes with $p=0$ and the optical angular momenta, $(\sigma,\ell)=( 1, 1 )$ and $(\sigma,\ell)=(-1, -1 )$, respectively. 
Basically, the squared magnitudes of the vector potentials of the TLs carrying finite OAM ($|\ell| > 0$) in the fundamental radial mode ($p=0$) present ring-shaped distributions over the $\vect{q}_{\parallel}$-plane, whose ring sizes increase with increasing $\ell$. \cite{peng2022tailoring} This indicates that the TLs with greater $\ell$ comprise the more components of large $q _{\parallel}$ and, according to the momentum-conservation law, likely couple the more exciton states with large in-plane momentum, $\vect{Q}$. Moreover, the effects of optical SOI become more important in the TLs with greater $\ell$. One also notes that the ring size of the $\vect{q}_{\parallel}$-dependent magnitudes of the longitudinal component, $\big| \mathcal{A}_{z}^{\sigma, \ell}( \vect{q}_{\parallel} ) \big| ^{2} $, is unequal but slightly larger than that of the transverse one, $\big| \mathcal{A}_{\parallel}^{\ell} (\vect{q}_{\parallel}) \big| ^{2}$. 
 With no effects of SOI, the {\it transverse} component of vector potential is decoupled from SAM (see Eq. (\ref{Aqp})) and remains the same for $\sigma=+1$ and $\sigma=-1$. Indeed, the patterns of Re${\left( \mathcal{A}_{\parallel}^{\ell=\pm1}\left( \vect{q}_{\parallel} \right)\right)}$ and Im${\left(\mathcal{A}_{\parallel}^{\ell=\pm1}\left( \vect{q}_{\parallel} \right)\right)}$ in Fig. S1(a.2)-(a.3) and S1(b.2)-(b.3) are shown dumbbell-like   to reflect the OAM $\ell=\pm1$ carried by the TL. 
As pointed out previously, the longitudinal field in a TL inherits the total angular momentum, $J=\sigma+\ell$, of the light.
Thus, as seen in Fig. S1(c.2)-(c.3) and S1(d.2)-(d.3), the in-plane patterns of the real and imaginary parts of $\mathcal{A}_{z}^{\sigma= \pm1 , \ell= \pm1}(\vect{q} _{\parallel})$  are double-dumbbell-like to reflect the TAM, $J=\sigma+\ell=\pm2$.

\begin{figure*}
	    \centering
        \includegraphics[width=\textwidth]{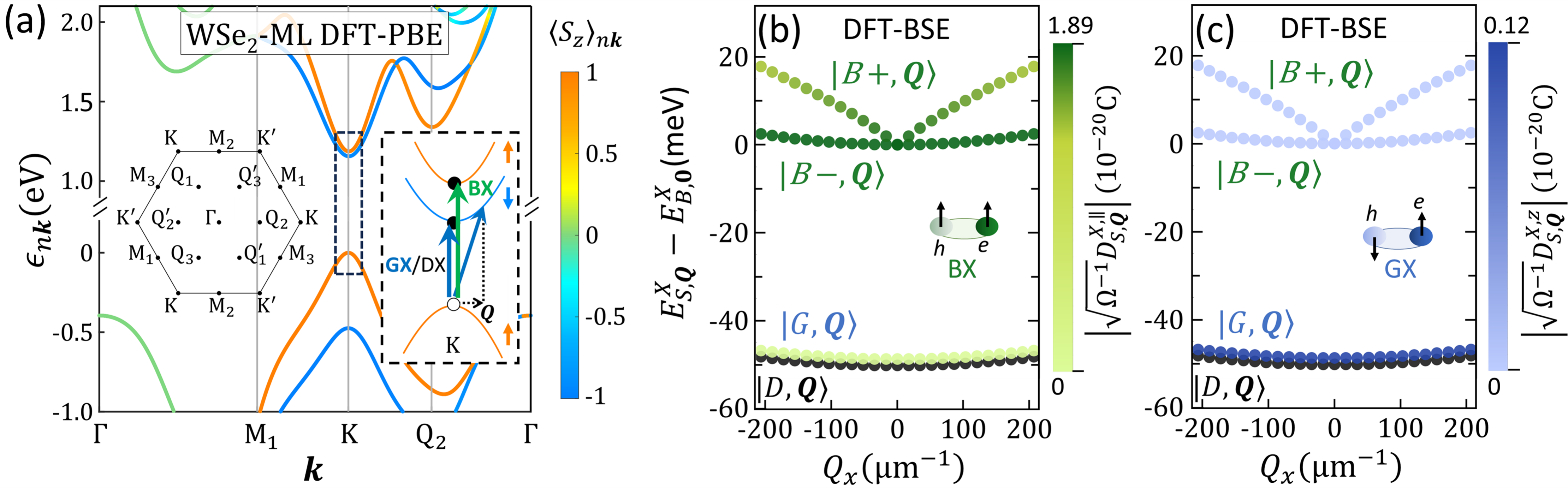}
	    \caption{(a) The spin-resolved quasi-particle band structure of a WSe$_{2}$-ML calculated by using the first-principles package Quantum Espresso with the PBE functional in DFT. The color of bands is mapped to the spin $z$-component of the Bloch states, $\langle S_{z} \rangle _{n \vect{k}}  = \langle \psi _{n \vect{k}} | \sigma _{z} | \psi _{n \vect{k}} \rangle$, where $\sigma _{z}$ is the Pauli matrix. The left inset shows the first Brillouin zone of TMD-MLs. The dashed rectangular box is the enlarged view of the lowest spin-split conduction bands and the topmost valence band at $K$-valley, where the green arrow line indicates the transitions of spin-allowed bright exciton (BX), the blue arrow line indicates the transition of spin-forbidden gray (GX) and dark exciton (DX), and the tilted arrow line indicates the transition of the exciton with a finite center-of-mass momentum, $\vect{Q}$. (b) The calculated 1s-exciton band structure, on the $Q_x$-axis, of a WSe$_2$-ML sandwiched by semi-infinite hBN layers obtained by solving the first-principles-based BSE, comprising the valley-split BX states, $|B\pm,\vect{Q}\rangle$, and the lowest GX and DX states, $|G,\vect{Q}\rangle$ and $|D,\vect{Q}\rangle$. The exciton bands are offset by the energy of BX at $\vect{Q} = \vect{0}$. The gradient green colors of the exciton states are mapped to the in-plane component of the transition dipole ($\vect{D} _{S,\vect{Q}} ^{X} = \vect{D} _{S,\vect{Q}} ^{X,\parallel} + D _{S,\vect{Q}} ^{X,z} \hat{\vect{z}}$) of exciton, $\big\vert \vect{D} _{S , \vect{Q}} ^{X , \parallel} \big\vert$. (c) shows the same exciton band structure as (b), but presents the out-of-plane component of the transition dipoles, $\big\vert D _{S , \vect{Q}} ^{X , z} \big\vert$, using gradient blue colors.}
	    \label{Fig2}
 \end{figure*}

\begin{figure*}
	    \centering
        \includegraphics[width=\textwidth]{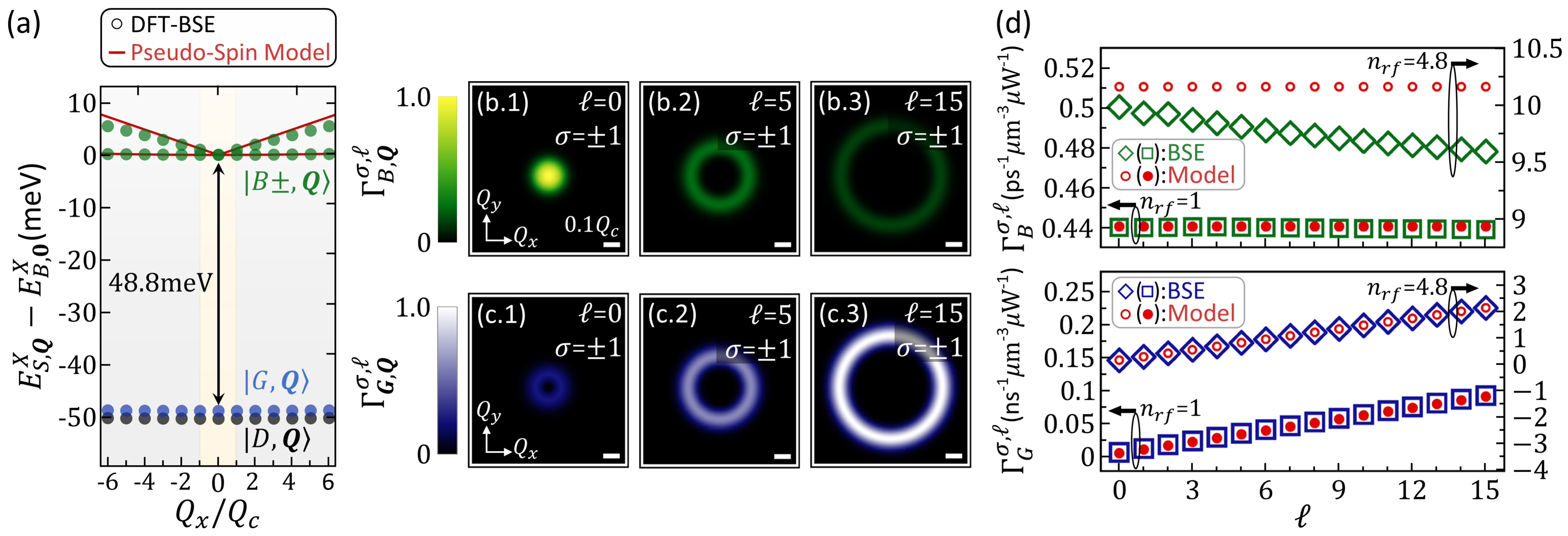}
	    \caption{ (a)  The exciton fine structure of the low-lying exciton states along the $Q_x$ axis around  the vicinity of the light-cone reciprocal range, comprising the valley-split bright exciton bands  (green circles) and the lowest gray (blue circles) and dark exciton ones (black circles).
	    $Q _{c} = E _{B \pm , \vect{0}} ^{X} / \hbar c$ is defined as the light-cone radius. For comparison, the band dispersions of BX states simulated by using the pseudo-spin model (red line) that consider a fixed value for the dipole of BX are presented.
	    (b1)-(b3) Density plots of the optical transition rates, $\Gamma_{B, \vect{Q}}^{\sigma , \ell} $ as functions of $\vect{Q}$ for the finite momentum BX states of a WSe$_2$-ML under the excitation of polarized TLs with the SAM and OAM, $(\sigma, \ell)= (\pm 1, 0), (\pm 1,5)$ and $(\pm 1, 15)$, respectively. (c1)-(c3) Density plots of $\Gamma_{G, \vect{Q}}^{\sigma , \ell} $ for the TL-excited finite-momentum GX states. All of the contour plots follow the same color-map on the leftmost side. For reference,  the length of the horizontal bar in white color represents the magnitude of 0.1$Q_c$. (d) The total transition rates of the TL-excited superposition states of BX (empty green squares) and GX states (empty blue squares) as a function of the $\ell$ of the TL, calculated by taking the exciton band structure obtained by solving the first-principles-based BSE from (a). For a more realistic simulation, the effective refractive index of TMD-MLs, \cite{PhysRevB.90.205422} $n _{rf} = 4.8$, that shortens the wavelength of light in materials ($q _{0} \rightarrow n _{rf} q _{0}$ and $Q _{c} \rightarrow n _{rf} Q _{c}$) is considered for comparison, as illustrated by the empty green and blue diamonds for BXs and GXs, respectively. In model simulations, where a fixed dipole value is assumed, the results for $n _{rf} = 1$ and $n _{rf} = 4.8$ are shown by the filled and empty red circles, respectively.}	     
	    \label{Fig3}
\end{figure*}

For the studies of exciton, we employ the theoretical methodology developed by Ref.\citen{peng2019distinctive,li2023key} to solve the Bethe-Salpeter equation (BSE) established in first-principles for the exciton fine structure spectra of encapsulated 2D materials (see Supporting Information for details).
The exciton states of a 2D material is expressed as $|S,\vect{Q}\rangle = \frac{1}{\sqrt{\Omega}} \sum _{v c \vect{k}} \Lambda _{S,\vect{Q}} (v c \vect{k}) \, \hat{c} _{c \vect{k} + \vect{Q}} ^{\dagger} \hat{h} _{v -\vect{k}} ^{\dagger} \left| GS \right\rangle$, where $\Omega$ is the area of the 2D material, $\hat{c} _{c \vect{k}} ^{\dagger}$ ($\hat{h} _{v -\vect{k}} ^{\dagger}$) is defined as the particle operator creating the electron (hole) of wavevector $\vect{k}$ ($- \vect{k}$) in conduction band $c$ (valence band $v$), $| GS \rangle$ denotes the ground state of the material, 
 $\Lambda _{S,\vect{Q}} (v c \vect{k})$ is the amplitude of the electron-hole configuration $\hat{c} _{c \vect{k} + \vect{Q}} ^{\dagger} \hat{h} _{v -\vect{k}} ^{\dagger} \left| GS \right\rangle$ and corresponds to the solution of the BSE for the exciton in momentum space, $S$ is the band index of the exciton state, $\vect{Q}$ is the center-of-mass momentum of exciton.
For a WSe$_2$-ML, the DX states as the exciton ground states are significantly lower than the bright ones by $\sim 48.8$ meV, as shown in Fig. \ref{Fig2}(b)-(c) and \ref{Fig3}(a). \cite{barbone2018charge,zinkiewicz2022effect} 
 The calculated BX-DX fine structure is in excellent agreement with the experimental observation. \cite{li2023key} 
Carefully examining the lowest DX states, one notes a small splitting between the DX doublet resulting. Combined with the SOI of quasi-particle, the inter-valley exchange interaction splits the lowest DX doublet and turns one of them, referred to as gray exciton (GX), to be slightly bright. \cite{robert2017fine} 
With finite $\vect{Q}$, the inter-valley {\it e-h} exchange interaction splits the valley exciton BX bands into a quasi-linear upper, $|B+,\vect{Q}\rangle$, and parabolic lower band, $|B-,\vect{Q}\rangle$. \cite{PhysRevB.91.075310,PhysRevLett.115.176801,yu2014dirac,simbulan2021selective,lin2023essential}
At the light cone edge where $\vert \vect{Q} \vert =q_0\approx Q_c$, the valley splitting between the upper and lower BX bands is merely  a few meV, much smaller than the energy separation of BX and DX/GX states. The transition dipole moment of an exciton state is evaluated by $\vect{D}^{X}_{S,\vect{Q}}=\frac{1}{\sqrt{\Omega}} \sum_{vc\vect{k}} \Lambda _{S,\vect{Q}} \left(v c \vect{k}\right) \vect{d}_{v \vect{k} , c \vect{k}}$, where $\vect{d} _{v \vect{k} , c \vect{k}} \equiv e \left\langle \psi _{v \vect{k}}\right\vert \vect{r} \left\vert \psi _{c \vect{k}} \right\rangle =  \frac{e\hbar}{im_0 (\epsilon _{v \vect{k}} - \epsilon _{c \vect{k}})} \langle \psi_{v \vect{k}} | \vect{p} | \psi_{c \vect{k}} \rangle$ is the dipole moment of single-electron transition evaluated by using the first-principles package Quantum Espresso and the Wannier90 package, as presented in Sec. SI.C of Supporting Information, \cite{kresse1996efficiency,kresse1996efficient,PhysRevLett.115.176801,peng2019distinctive}  with $\vect{p}$  the operator of linear momentum, $m _{0}$ ($|e|$) the electron rest mass (the elementary charge), and $\epsilon _{n \vect{k}}$ the energy of the Bloch state characterized by the band index $n$ and wavevector $\vect{k}$. The first-principles-calculated quasi-particle band structure of a WSe$_{2}$-ML is shown in Fig. \ref{Fig2}(a).
The first-principles-calculated $\vect{Q}$-dependent in-plane and out-of-plane projections of the transition dipole of the exciton states in the fine structure of WSe$_{2}$-ML are shown in Fig. \ref{Fig2}(b) and \ref{Fig2}(c), respectively.
 
The numerically calculated transition dipole moments of the BX (GX) states are shown mainly in-plane (out-of-plane) oriented,  as presented by Fig. \ref{Fig2}(b) and (c).
In addition to the strong exciton-photon interaction, the fine structure spectrum of a WSe$_2$-ML consisting of various exciton states with distinctly oriented dipoles serves as an excellent test bed to explore the distinct field components in the 3D-structured lights. In turn, TLs carrying controlled SAM and OAM enable us to selectively access and distinguish a variety of exciton states of 2D materials.

In the widely-used exciton pseudo-spin model that neglects the slight variations of dipole moments with respect to $\vect{Q}$, the transition dipoles of the upper BX, lower BX, and GX states are described by $\vect{D} _{B + , \vect{Q}} ^{X} \approx D _{B + , \vect{0}} ^{X} \hat{\vect{Q}}$, $\vect{D} _{B - , \vect{Q}} ^{X} \approx D _{B - , \vect{0}} ^{X} \hat{\vect{Q}} _{\perp}$, and $\vect{D} _{G , \vect{Q}} ^{X} \approx D _{G , \vect{0}} ^{X} \hat{\vect{z}}$, respectively, where $\hat{\vect{Q}} = \vect{Q} / | \vect{Q} |$ , $\hat{\vect{Q}} _{\perp} = - \sin \phi _{\vect{Q}} \hat{\vect{x}} + \cos \phi _{\vect{Q}} \hat{\vect{y}}$ ($\hat{\vect{Q}} _{\perp} \cdot \hat{\vect{Q}} = \hat{\vect{Q}} _{\perp} \cdot \hat{\vect{z}} = 0$), and $D _{B \pm , \vect{Q}} ^{X}$ ($D _{G , \vect{Q}} ^{X}$) is the magnitude of the dipole moment of BX (GX). With the fixed magnitude of the exciton dipole, the energy band dispersion of BX doublet split by the long-ranged exchange interaction can be explicitly solved and shown in Fig. \ref{Fig3}(a) in comparison with the first-principles-calculated results.

In the time-dependent perturbation theory, the Hamiltonian of light-matter interaction with respect to a light described by  the vector potential $\vect{A}(\vect{r})$ is given by $H_{LMI}\approx \frac{|e|}{2m_0}  \vect{A}(\vect{r})\cdot \vect{p} $ in 
the weak field and rotating wave approximations. \cite{parker2018physics}
The first-order perturbed exciton state could be expressed as \cite{peng2022tailoring}
\begin{align}
	| \Psi ^{X} \rangle \approx | GS \rangle + \sum _{S \vect{Q}} \tilde{c} _{S , \vect{Q}} e ^{-i E _{S , \vect{Q}} ^{X} t / \hbar} | S , \vect{Q} \rangle \label{EWP} \, ,
\end{align}
where $\tilde{c} _{S , \vect{Q}} = \tilde{c} _{S , \vect{Q}} (t)$ is the time-dependent coefficient of the exciton state, $| S , \vect{Q} \rangle$.
As shown in Eq. (\ref{EWP}), the exciton state photo-excited by a structure light will be in the exciton superposition state, which is superimposed by the states of individual finite-momentum exciton, $| S , \vect{Q} \rangle$, with the complex coefficient that is shown to be proportional to the optical matrix element $\tilde{c} _{S , \vect{Q}} \propto \tilde{M} _{S , \vect{Q}}$. The optical matrix element of an exciton state, $\vert S,\vect{Q} \rangle $, is derived as
$\tilde{M}_{S,\vect{Q}}^{\sigma, \ell} = \frac{1}{\sqrt{\Omega}}\sum_{vc\vect{k}} \Lambda^* _{S,\vect{Q}} (v c \vect{k}) \langle \psi_{c\vect{k} +\vect{Q}}|\frac{|e|}{2m_0} \vect{A} ^{\sigma , \ell} (\vect{r})\cdot \vect{p}|\psi_{v\vect{k}}\rangle$, \cite{peng2022tailoring} 
which measures the amplitude of the optical transition of the exciton state, $\vert S,\vect{Q}\rangle$, induced by  the incident TL carrying the angular momenta $\sigma$ and $\ell$. 
In terms of the optical matrix element, the Fermi's golden rule formulates the rate of incoherently photo-exciting the finite-momentum exciton state, $|S,\vect{Q}\rangle $, by using a TL with $(\sigma,\ell)$ as 
\begin{align}
	\Gamma _{S,\vect{Q}} ^{\sigma , \ell} =\frac{2\pi}{\hbar} \vert \tilde{M}_{S,\vect{Q}}^{\sigma, \ell} \vert^2 \rho(\hbar\omega=E_{S,\vect{Q}}^X) \label{FermiGR} ,
\end{align}
where $\rho(\hbar\omega)$ is the density of states  of light in the range of angular frequency between $\omega$ and $\omega+d\omega$. 
 The incoherence nature of the Fermi’s golden rule can be better understood in the density matrix representation of wavefunction, which defines the density matrix element $\rho _{S , \vect{Q} ; S ^{\prime} , \vect{Q} ^{\prime}} \equiv \tilde{c} _{S , \vect{Q}} \tilde{c} _{S ^{\prime} , \vect{Q} ^{\prime}} ^{*}$. \cite{blum2012density} Thus, in the incoherent time-dependent perturbation theory, the transition rate of the exciton superposition state can be counted by summing all the transition probabilities of $| S , \vect{Q} \rangle$, which is measured by the real-valued diagonal element of the density matrix for the superposition states, i.e. $\rho _{S , \vect{Q} ; S , \vect{Q}} = | \tilde{c} _{S , \vect{Q}} | ^{2} \propto | \tilde{M} _{S , \vect{Q}} | ^{2}$.
 
In the electric dipole approximation, one derives
\begin{align}
\tilde{M}_{S,\vect{Q}}^{\sigma, \ell} \approx \frac{E_g}{2i\hbar} \vect{\mathcal{A}}^{\sigma, \ell} (\vect{Q}) \cdot \vect{D}_{S,\vect{Q}}^{X\ast} \label{Ms} \, ,
\end{align}
where $\vect{\mathcal{A}}^{\sigma, \ell} (\vect{Q}) = \hat{\vect{\varepsilon}}_{\parallel}^{\sigma} \mathcal{A}_{\parallel}^{\ell}(\vect{Q}) + \hat{\vect{\varepsilon}} _{z}  \mathcal{A}_{z}^{\sigma, \ell} (\vect{Q})$ is the Fourier transform of the vector potential of structured light with the transverse and longitudinal components as given by Eqs. (\ref{Aqp}) and (\ref{Aqz}), respectively, $E_g=\epsilon _{c_1 \vect{K}/\vect{K}'} - \epsilon _{v_1 \vect{K}/\vect{K}'}$ is the energy gap of the material at the $\vect{K}/\vect{K}'$ point, and $c_1$ ($v_1$) is the lowest conduction (topmost valence) band. The interaction between TL and excitons is characterized by the optical matrix element of Eq. (\ref{Ms}). This element involves the inner product of the optical polarization vector, $\hat{\vect{\varepsilon}} _{\parallel} ^{\sigma}$ and $\hat{\vect{\varepsilon}} _{z}$, and the exciton dipole, $\vect{D} _{S , \vect{Q}} ^{X}$, which is further weighted by the $\vect{Q}$-dependent scalar functions, $\mathcal{A} _{\parallel} ^{\ell} (\vect{Q})$ and $\mathcal{A} _{z} ^{\sigma , \ell} (\vect{Q})$. Basically, the inner product $\hat{\vect{\varepsilon}} \cdot \vect{D} _{S , \vect{Q}} ^{X \, *}$ (with $\hat{\vect{\varepsilon}} = \hat{\vect{\varepsilon}} _{\parallel} ^{\sigma}$ or $\hat{\vect{\varepsilon}} = \hat{\vect{\varepsilon}} _{z}$) in Eq. (\ref{Ms}) remains the same form for both TLs and regular non-structured lights. However, the scalar functions $\mathcal{A} _{\parallel} ^{\ell} (\vect{Q})$ and $\mathcal{A} _{z} ^{\sigma , \ell} (\vect{Q})$, which reflect the spatial structures of TLs, will impose a weighting on the optical matrix element for a finite-momentum exciton state, $| S , \vect{Q} \rangle$. This indicates that under the electric dipole approximation, the optical selection rules for a specific state $| S , \vect{Q} \rangle$ that are generally determined by $\hat{\vect{\varepsilon}} \cdot \vect{D} _{S , \vect{Q}} ^{X \, *}$ remain unchanged for both TLs and regular non-structured lights, but the spatial structures of TLs will directly affect the weighting of the exciton state component, $| S , \vect{Q} \rangle$, with the center-of-mass momentum $\vect{Q}$ in the exciton superposition state (see Eq. (\ref{EWP})). In other words, the spin part (polarization) of the light directly couples with the dipole of exciton that is related to the relative coordinate of exciton, while the orbital part of the light couples the center-of-mass movement of the exciton. \cite{PhysRevLett.89.143601}

Deriving optical selection rules for exciton states under TL excitation is not a trivial task. The complexity arises because TL-excited superposition states, as detailed in Eq. (\ref{EWP}), involve finite-momentum exciton states that do not align with high-symmetry points in reciprocal space. The pioneering work in Ref. \citen{PhysRevB.108.125435} analyzed the optical selection rules for BX and GX states in 2D materials under single TL excitation, but applied its group theory analysis only to zero-momentum exciton states. In contrast, our work inherently includes the complete symmetries of finite-momentum exciton states by considering the crystal symmetries within the crystal structures during our first-principles computations.

The optical matrix elements of Eq. (\ref{Ms}) for BX and GX states under the excitation of a TL in the LG mode with $(\sigma, \ell)$ are derived in the cylindrical coordinate and explicitly shown as below, 
\begin{align}
 \tilde{M}_{B\pm, \vect{Q}}^{\sigma, \ell} \approx (i \sigma) ^{(1 \mp 1)/2} \frac{E_g }{2\sqrt{2}i\hbar} \tilde{F}_{|\ell| } \left( Q \right) D _{B \pm , \vect{Q}} ^{X} e^{i (\sigma + \ell) \phi_{\vect{Q}}} \label{M_B}
\end{align} 
and
\begin{align}
\tilde{M}_{G, \vect{Q}}^{\sigma, \ell} \approx  - \frac{E_g }{2\sqrt{2}i\hbar} \tilde{F}_{|\ell| } \left( Q \right) D _{G , \vect{Q}} ^{X} e^{i (\sigma + \ell) \phi_{\vect{Q}}} \sin\theta_{\vect{Q}}\, , \label{M_G}
\end{align}
where the exponential term $e^{i (\sigma + \ell) \phi_{\vect{Q}}}$ accounts for the TAM transfer from a TL to a GX and the term  $\sin\theta_{\vect{Q}} \equiv \frac{Q}{\sqrt{Q^2+ q_0^2}}$ arises from the SOI, which makes a normally incident TL forbidden to excite a GX with $Q=0$ but enhances the photo-generate of GX states with large $Q$ as increasing $\ell$. 

The exciton fine structure splitting in 2D materials should be an essential material property in the transfer of optical information encoded in a VVB. With the relatively large splitting between BX and GX states, \cite{PhysRevB.93.121107} the GX states of a W-based TMD-ML are particularly advantageous for receiving the optical information carried by VVBs since they are unlikely to mix with the high-lying BX states to mess up the received optical information. In contrast, in Mo-based TMD-MLs, the energy levels of the GX states are usually very close to those of dipole-allowed BX states, typically with a small energy separation of a few meV. \cite{robert2020measurement,lu2019magnetic} As compared with W-based TMD-MLs, the small energy separation between BX and GX states of Mo-based TMD-MLs might set an uncertainty in the preservation of the optical information imprinted onto GX states in 2D materials.

Since the valley splitting between the lower and upper BX bands is merely a few meV and normally spectrally unresolvable, as seen in Fig. \ref{Fig3}(a), \cite{schneider2020optical} the total transition rate of the BX doublet, $\vert B\pm, \vect{Q}\rangle$, under the photo-excitation of a TL can be counted by $ \Gamma_{B,\vect{Q}}^{\sigma, \ell} \equiv  \sum_{S=B\pm} \Gamma_{S, \vect{Q}}^{\sigma, \ell} \propto \sum_{S=B\pm} \vert \tilde{M}_{S,\vect{Q}}^{\sigma, \ell}\vert^2  $. 
By contrast, a GX state is normally spectrally well apart from the BX states and its transition rate is evaluated by
 $\Gamma_{G, \vect{Q}}^{\sigma, \ell} \propto  \vert \tilde{M}_{G, \vect{Q}}^{\sigma, \ell}\vert^2 $.

Figure \ref{Fig3}(b) and (c) show the density plots of the optical transition rates, $\Gamma _{B, \vect{Q}}^{\sigma, \ell}$ and $\Gamma _{G, \vect{Q}}^{\sigma, \ell}$, as functions of $\vect{Q}$ for the finite-momentum BX and GX states of a WSe$_2$-ML incident by polarized TLs with $(\sigma, \ell) = (\pm 1, 0)$, $(\pm 1, 5)$ and $(\pm 1, 15)$, respectively. Overall, the $ \Gamma_{B/G, \vect{Q}}^{\sigma, \ell}$ for the non-zero $\ell = 5$, 15 exhibit similar ring-shaped patterns over the $\vect{Q}$-plane, with the  ring sizes increasing with increasing $\ell$.
This indicates that a TL with greater $\ell$ enables the photo-generation of the exciton states (both BX and GX ones) with larger $Q$, whose superposition forms a  more localized spatially wave packet as previously pointed out by Ref.  \citen{peng2022tailoring}. 

For a comprehensive understanding, it is interesting to figure out how BX and GX states respond differently when excited by the TL with a well-defined OAM but vanishing SAM (such as linear polarization) and the TL with zero OAM and a well-defined SAM (circular polarization). The simulation results for this part of the discussion are shown in Fig. S3 of the Supporting Information. For BXs, both types of light result in an isotropic pattern in the transition rates, as depicted in the upper panels of Fig. S3. In contrast, GXs exhibit a notable difference. The lower panels of Fig. S3 show that the pattern of the transition rates for GXs becomes anisotropic when excited by light with well-defined OAM but vanishing SAM, whereas it remains isotropic when excited by light with zero OAM and well-defined SAM. This feature, as seen in the lower panel of Fig. S3(a), can be realized by examining Eq. (\ref{Aqz}), where the $\vect{q} _{\parallel}$-dependent vector potential of the longitudinal field is directly determined by the inner product of $\hat{\vect{\varepsilon}}$ and $\hat{\vect{q}}$. In the case of linearly polarized TL, this results in anisotropy in the transition rate pattern of GXs. This clearly demonstrates the distinct effects of polarization (SAM) and spatial vortex (OAM) in structured lights on GXs.

Analytically, one can show that the TL with $\ell$ mostly likely excite the finite momentum BX state with $Q = q_{\parallel}^{\ell} = \sqrt{2 | \ell|}/w_0$,  where the squared magnitude of ${\mathcal{A}_{\parallel}^{\ell}(\vect{Q})}$ is maxima so that $\partial | \mathcal{A} _{\parallel} ^{\ell} (\vect{Q}) |^{2} / \partial Q | _{Q = q _{\parallel} ^{\ell}} = 0$. The upper (lower) panel of Fig. \ref{Fig3}(d) shows the total transition rate, $\Gamma_{B/G}^{\sigma, \ell} \equiv \sum_{\vect{Q}} \Gamma_{B/G, \vect{Q}}^{\sigma, \ell}$ as a function of $\ell$, of BX (GX), summing up the contributions by all the finite-momentum states excited by the TLs with $\ell=0,1,...15$. 
Notably, the rate of photo-exciting the GX superposition states, $\Gamma_{G}^{\sigma, \ell} $, using a TL with $\ell$ is shown linearly increasing with increasing $\ell$, while the rate of photo-exciting the BX ones, $\Gamma_{B}^{\sigma, \ell}$, remain nearly unchanged against $\ell$. Increasing the OAM of the incident TL from $\ell=1$ to $\ell=15$, $\Gamma_{G}^{\sigma, \ell}$ is enhanced by over one order of magnitude. The $\ell$-enhanced photo-generation of GX is associated with the term of SOI, $(\hat{\vect{\varepsilon}}_{\parallel}^{\sigma} \cdot \hat{\vect{{q}}}) = \frac{q_{\parallel}}{\sqrt{2(q_{\parallel}^2 +q_0^2}) } e^{i \sigma \phi_{\vect{q}}} \approx \frac{1}{\sqrt{2}} \frac{q_{\parallel}}{q_0} e^{i \sigma \phi_{\vect{q}}} \propto q_{\parallel}$, in the longitudinal field of TL as expressed by  Eq. (\ref{Aqz}). 
From $\partial | \mathcal{A} _{z} ^{\sigma , \ell} (\vect{Q}) | ^{2} / \partial Q | _{Q = \bar{q} _{\parallel} ^{\ell}} = 0$, we have $Q = \bar{q} _{\parallel} ^{\ell} = \sqrt{2(| \ell | + 1)} / w _{0}$.
Thus, with increasing $\ell$ of a TL, the in-plane component of momentum, $q_{\parallel}$, carried by the TL increases, and so do the strength of the optical SOI and the magnitude of the longitudinal field, ${\mathcal{A} _{z} ^{\sigma, \ell} (\vect{Q})}$, of Eq. (\ref{Aqz}).
In Fig. \ref{Fig3}(d), alongside the first-principles calculated results, we also present the model-simulated transition rates (indicated by the filled/empty red circles). Due to the weak momentum dependence of the optical matrix element within the narrow reciprocal area of the light-cone, the first-principles-calculated transition rate of the BX (GX) is slightly lower than (very similar to) the model-simulated ones.

Despite the phase term of TAM, $e^{i(\sigma+\ell)\phi_{\vect{Q}}}$, encoded in the complex optical matrix elements of BX and GX states  as shown in  Eqs. (\ref{M_B}) and  (\ref{M_G}), the phase information is not preserved in the squared magnitude of the optical matrix elements. These squared magnitudes, which measure the optical transition rates of the exciton states under incoherence conditions, cannot show the transferred angular momenta to the exciton states. 
Numerically, by using the incoherent theory of Fermi’s golden rule, we will show that the optical information encoded in the longitudinal field of a VVB at the equator state of Poincaré sphere is still transferable to the GX states. This is because, before the photo-generation of an exciton, the internal interference between two TL components in a VVB occurs in advance and directly encodes the optical information into the real-valued amplitude, rather than the complex phase term, of light. In the manner, the read-out of the imprinted optical information in the GX states by a VVB can be achieved by using the regular optical spectroscopy with angle-resolution, as shown in the later section and Fig. \ref{Fig6}.

\begin{figure*}
	    \centering
	    \includegraphics[width=\textwidth]{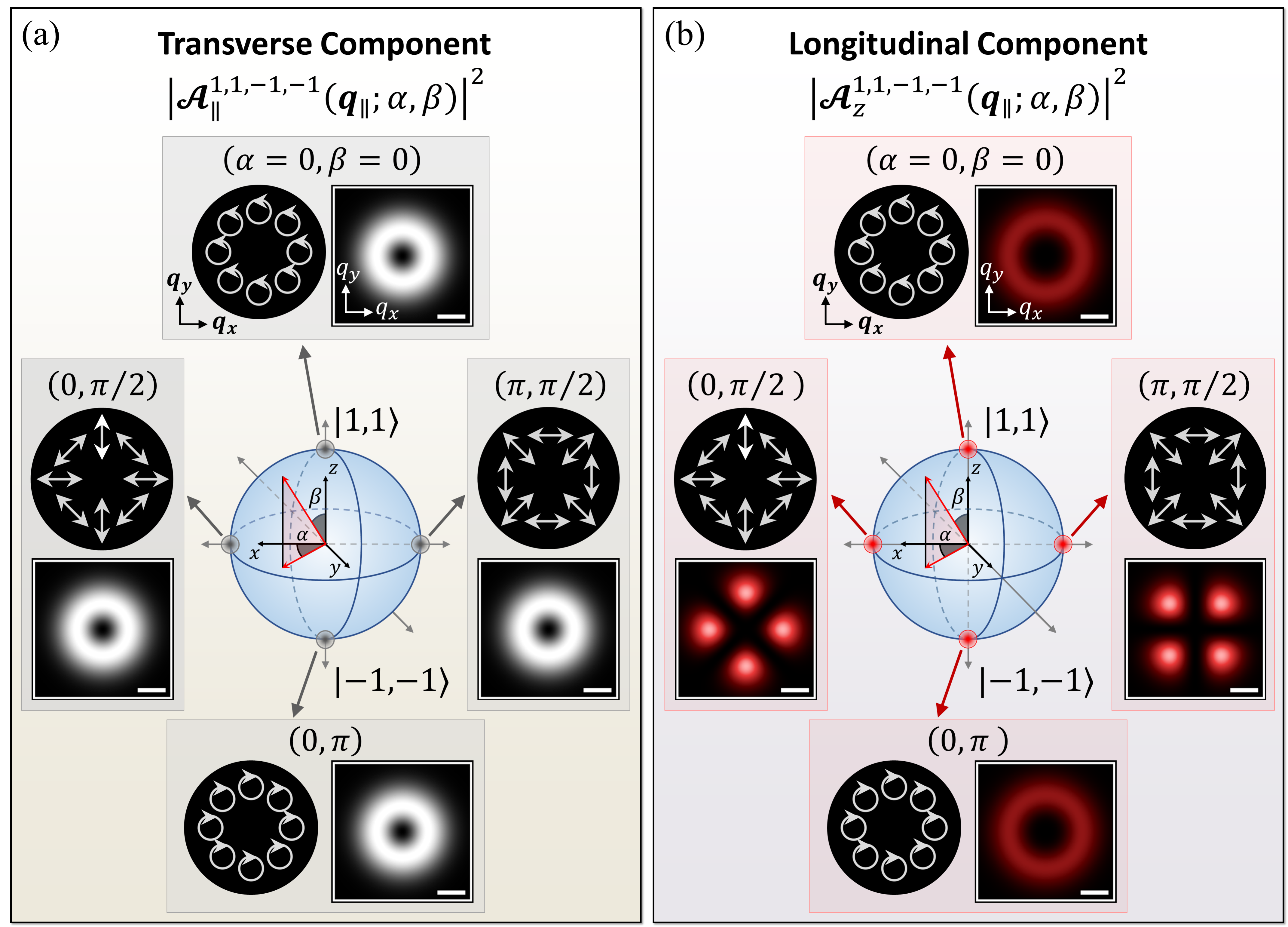}
	    \caption{Representation of higher-order Poincaré sphere for the superposition states, $\vert \sigma , \ell,\sigma' , \ell' ; \alpha , \beta \rangle $, of VVBs in the basis of the two TLs with $\left(\sigma=1,\ell=1\right)$ and $\left(\sigma'=-1,\ell'=-1\right)$. (a) Density plots of the squared magnitudes of the transverse  components of the vector potentials, $\big| \mathcal{A}_{\parallel}^{1,1,-1,-1} (\vect{q}_{\parallel};\alpha,\beta) \big| ^{2}$, for the VVB states at the poles and equator of  higher-order Poincaré sphere over the $\vect{q} _{\parallel}$-plane. For reference, the length of the horizontal bar in white color represents the magnitude of $q_{\parallel}= 0.1Q_c$.  (b) Density plots of the squared magnitudes of the longitudinal components of the vector potentials, $\big| \mathcal{A}_{z}^{1,1,-1,-1} (\vect{q}_{\parallel};\alpha,\beta ) \big| ^{2}$, of the same VVBs as presented in (a).  While the pattern of  $\big| \mathcal{A}_{\parallel}^{1,1,-1,-1} (\vect{q}_{\parallel};\alpha,\beta) \big| ^{2}$ always remain isotropic as varying the geometric angles of the superposition states of VVB, those of  $\big| \mathcal{A}_{z}^{1,1,-1,-1} (\vect{q}_{\parallel};\alpha,\beta=\pi/2) \big| ^{2}$ of the equatorial superposition states exhibit anisotropic patterns, possessing the rotational symmetry associated with the finite $\ell\neq 0$ carried by the TL basis of the VVB. The dark circular panels in (a) and (b) present the spatially varying transverse polarizations of the VVBs over the $\vect{q} _{\parallel}$-plane. The pure states of TL basis at the poles $\left( \beta=0,\pi \right)$ possess circular polarization. By contrast, the maximal superposition states of VVB at the equatorial points $\left(\beta=\pi/2 \right)$,  are linearly polarized along the direction depending on $\vect{q}_{\parallel}$.}
	     \label{Fig4}
\end{figure*}

Superimposing two TLs with distinct SAM ($\sigma'=-\sigma$) and OAM ($\ell\neq \ell'$) forms a VVB, $|\sigma , \ell, \sigma' , \ell' ; \alpha , \beta \rangle$, \cite{ mclaren2015measuring,shen2019optical,Forbes2021structured}  which is structured in both polarization and amplitudes over the 3D space, \cite{zhao2023generation} 
and is prospective in the frontier photonic applications, \cite{rosales2018review} e.g. laser material processes, \cite{weber2011effects,toyoda2013transfer} optical encoding/decoding in communication, \cite{zhao2015high} and microscopy. \cite{chen2013imaging}
In Fig. \ref{Fig4}, the north pole $(\alpha = 0 , \beta = 0)$ and south pole $(\alpha = 0 , \beta = \pi)$ represent the TL basis in the single LG modes, which are $|\sigma , \ell \rangle$ and $|\sigma' , \ell'\rangle$, respectively. The superposition state, $|\sigma , \ell, \sigma' , \ell' ; \alpha , \beta \rangle$, with $\beta \neq 0$ or $\pi$ is represented by points located on the sphere surface in between the poles. In this manner, the maximal superposition state is positioned at the equator ($\beta = \pi /2$), in contrast to the north/south pole, which no longer signifies a VVB but instead represents an individual TL state.

The vector potential of such a VVB is given by
$\vect{ \mathcal{A} }^{\sigma,\ell, -\sigma,\ell'}(\vect{q}_\parallel;\alpha,\beta)=\cos \left( \beta/2 \right) \vect{\mathcal{A}}^{\sigma, \ell}(\vect{q}_\parallel) + e^{i\alpha} \sin \left( \beta/2 \right) \vect{\mathcal{A}}^{-\sigma,\ell'}(\vect{q}_\parallel)$  and leads to the corresponding complex optical matrix element for an exciton in the  state  $\vert S,\vect{Q} \rangle $, $\tilde{M}_{S,\vect{Q}}^{\sigma, \ell, -\sigma, \ell'}(\alpha,\beta) = \cos \left( \beta/2 \right) \tilde{M}_{S,\vect{Q}}^{\sigma, \ell} + e^{i\alpha} \sin \left( \beta/2 \right) \tilde{M}_{S,\vect{Q}}^{-\sigma, \ell'}$.
Following Eq. (\ref{Aqp}), one can show that the squared magnitude of the transverse component of the vector potential in angular spectrum representation is 
 $\big| \vect{\mathcal{A}}_{\parallel}^{\sigma,\ell \, , -\sigma, \ell'} (\vect{q}_\parallel; \alpha, \beta)\big| ^{2} = \cos ^{2} (\beta / 2) F _{|\ell|} ^{2} (q _{\parallel}) +\sin ^{2} (\beta / 2) F _{|\ell'|} ^{2} (q _{\parallel})$.
 One notes that the amplitude of the in-plane field of VVB is independent of the azimuthal angle, $\phi _{\vect{q}}$.  Thus, $\big| \vect{\mathcal{A}}_{\parallel}^{\sigma,\ell \, , -\sigma, \ell'} (\vect{q}_\parallel; \alpha, \beta) \big| ^{2}$ exhibits the isotropic contours over the $\vect{\vect{q}}_{\parallel}$-plane, as shown in Fig. \ref{Fig4}(a), and does not preserve the optical information of $\ell$ carried by the TL basis that is encoded in the phase term, $e^{i\ell\phi_{\vect{q}}}$, of Eq. (\ref{Aqp}).

By contrast, the squared magnitude of the longitudinal component of the vector potential of the same VVB is derived as 
$\big\vert \vect{\mathcal{A}}_{z}^{\sigma,\ell , -\sigma, \ell'} (\vect{q}_\parallel; \alpha, \beta) \big\vert ^{2} = \frac{\sin ^{2} \theta_{\vect{q}}}{2} [\cos ^{2} \frac{\beta}{2} F _{|\ell|} ^{2} \left(q_{\parallel} \right) + \sin ^{2} \frac{\beta}{2} F _{|\ell'|} ^{2} \left(q_{\parallel} \right) \,\,\,\,\,\,\,\,\,\,\,\, +\,\,\,\,\,\,\,\,\,\,\,\, \sin \beta \, F_{|\ell|}\left(q_{\parallel} \right) F _{|\ell'|} \left( q _{\parallel} \right) \\ \cos \left(\Delta  J \,\phi_{\vect{q}}+\left(\alpha-  \frac{\pi}{2} (\vert \ell'\vert - \vert \ell \vert) \right) \right)] $ and shown the $\phi_{\vect{q}}$-dependence, as long as $\beta\neq 0, \pi$ and $\Delta J \equiv J'-J = (\sigma'+\ell')-(\sigma+\ell) \neq 0$.

Interestingly, \, the\, $\phi_{\vect{q}}$-dependence\, of 
$\big\vert \vect{\mathcal{A}}_{z}^{\sigma, \ell , -\sigma, \ell'} (\vect{q}_\parallel; \alpha, \beta) \big\vert ^{2} $ is featured with the winding number $\Delta J =\ell'-\ell -2\sigma$ that reflects the difference of TAM between the two TL basis. 
Figure \ref{Fig4}(b) shows the distribution of the squared magnitude of the longitudinal component of the vector potential, $\big\vert \vect{\mathcal{A}}_{z}^{1,\, 1 , -1,\, -1} (\vect{q}_\parallel; \alpha, \beta) \big\vert ^{2}$, over the $\vect{q} _{\parallel}$-plane for the VVB superposed by the TLs with $(\sigma , \ell)=(1 , 1)$ and $(\sigma' , \ell')=(-1 , -1)$.
Indeed, we observe the anisotropic patterns of $\big\vert \vect{\mathcal{A}}_{z}^{1,\, 1 , -1,\, -1} (\vect{q}_\parallel; \alpha, \beta=\pi/2) \big\vert ^{2}$ in the four-fold rotational symmetry, matching $\Delta J = -4 $ of the VVB.

\begin{figure*}
	    \centering
        \includegraphics[width=\textwidth]{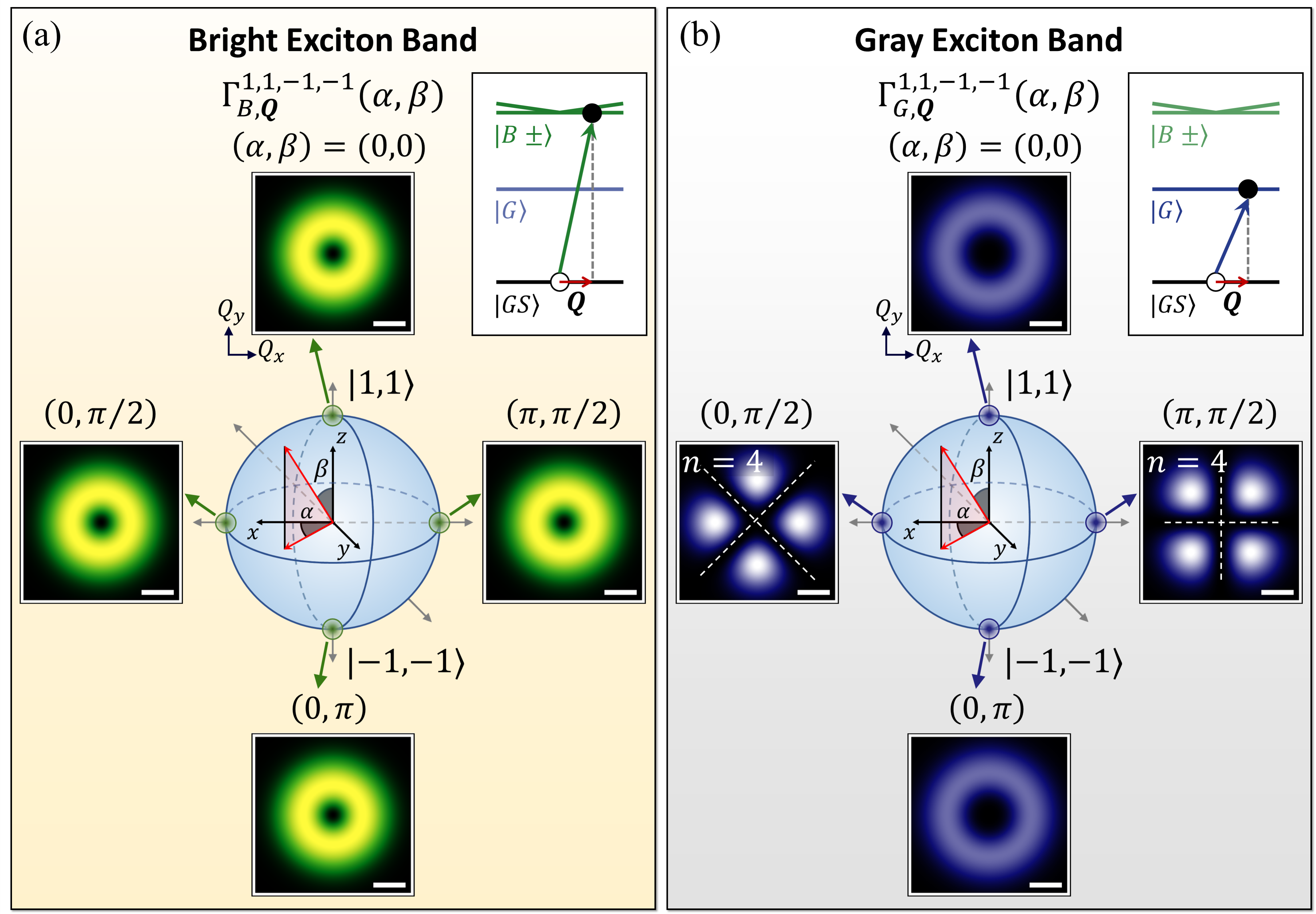}
	    \caption{ (a) Density plots of the  $\vect{Q}$-dependent transition rates,  $\Gamma_{B,\vect{Q}}^{1, 1, -1, -1} (\alpha , \beta) $, of the finite-momentum BX doublet,  $\vert B\pm,\vect{Q} \rangle$, and (b) the  $\vect{Q}$-dependent transition rates, $\Gamma_{G,\vect{Q}}^{1, 1, -1, -1} (\alpha , \beta)$, of the finite-momentum GX state, $\vert G,\vect{Q} \rangle$, under the excitation of the VVBs in the TL-superposition states,  $\vert 1, 1 , -1,-1; \alpha , \beta \rangle$, with the different geometric angles,  $(\alpha,\beta)=(0,0)$, $(0, \pi)$, $(0, \pi/2)$ and $(\pi, \pi/2)$, in the higher order Poincar\'e sphere representation. The schematic inset in each panel illustrates the optical transition corresponding to each case.  For reference, the length of the horizontal bar in white color represents the magnitude of $Q= 0.1Q_c$. Notably, $\Gamma_{G,\vect{Q}}^{1, 1, -1, -1} (\alpha ,\beta=\pi/2)$ exhibits the anisotropic patterns with the four-fold rotation symmetry ($n=4$), from which the angular momentum, $\ell$  and $\ell' \, (= - \ell)$, carried by the TL basis can be inferred according to Eq. (\ref{ell}).}
	    \label{Fig5}
\end{figure*}

Further, from Eq. (\ref{M_B}) one can derive the total transition rate of the spectrally unresolvable BX doublet with $\vect{Q}$ under the excitation of a VVB, $ \Gamma_{B, \vect{Q}}^{\sigma,\ell \, , -\sigma, \ell'}  \propto \sum_{S = B \pm} \big\vert \tilde{M}_{S,\vect{Q}}^{\sigma,\ell \, , -\sigma, \ell'} \big\vert ^{2} = \left[ \frac{E _{g}}{2 \hbar} \left( \frac{D _{B + , \vect{Q}} ^{X} + D _{B - , \vect{Q}} ^{X}}{2} \right) \right] ^{2} \big[ \cos ^{2} \frac{\beta}{2} F _{|\ell|} ^{2} (Q) + \sin ^{2} \frac{\beta}{2} \\ F _{|\ell'|} ^{2} (Q) \big]$.  As expected, the transition rate of BX doublet, $ \Gamma_{B, \vect{Q}}^{\sigma,\ell \, , -\sigma, \ell'}$, excited by a VVB is shown $\phi_{\vect{Q}}$-irrelevant and exhibit an isotropic distribution over the $\vect{Q}$-plane, as shown by Fig. \ref{Fig5}(a) for the VVB with $(\sigma , \ell)=(1 , 1)$ and $(\sigma' , \ell')=(-1 , -1)$.   
For a GX, the transition rate, $\Gamma_{G,\vect{Q}}^{\sigma, \ell, \sigma', \ell'}(\alpha,\beta) \propto \big\vert M_{G,\vect{Q}}^{\sigma, \ell, \sigma', \ell'}(\alpha,\beta) \big\vert ^{2} $, is derived as

\begin{strip}
\begin{equation}
 \resizebox{1\linewidth}{!}{
$\Gamma_{G,\vect{Q}}^{\sigma, \ell, \sigma', \ell'}(\alpha,\beta) \propto \left(\frac{E_g D ^{X}_{G , \vect{Q}}}{2\hbar}\right)^2\frac{\sin^2\theta_{\vect{Q}}}{2} \left[ \cos^2\frac{\beta}{2} F _{|\ell|} ^{2} (Q)  +\sin^2\frac{\beta}{2} F _{|\ell'|} ^{2}(Q) + F_{|\ell|}(Q)F_{|\ell'|}(Q)\sin\beta \cos\left(\Delta J\,\phi_{\vect{Q}}+\left(\alpha-\Delta|\ell| \frac{\pi}{2}\right)\right)\right]\, , $} 
 \label{TGX}
\end{equation}
\end{strip}

where $\Delta \vert \ell\vert \equiv \vert \ell' \vert - \vert \ell \vert $. The first two terms in Eq. (\ref{TGX}) can be viewed as the sum of the squared magnitude of the transition rate of GX under the excitation of the two non-interfered TL-basis of the VVB, which depends only on the magnitude of $\vect{Q}$ and remains invariant with varying $\phi_{\vect{Q}}$. The last cross-term arises from the coherent interference between the two TL basis and explicitly shows the $\phi_{\vect{Q}}$-dependence, which is importantly associated with the difference of TAM between the TL basis, $\Delta J$. As $\left(\alpha-\Delta\vert\ell\vert\pi/2\right) $ is simply a  constant phase, the last cross-term of Eq. (\ref{TGX}),  $\propto \cos\left(\Delta J\,\phi_{\vect{Q}}+(\alpha-\Delta|\ell| \frac{\pi}{2})\right)$,  varies sinusoidally with the winding number, i.e. $n=|\Delta J|$, by rotating $\phi_{\vect{Q}}$.

This suggests that one can decode the angular momentum difference ($\Delta J$) within the VVB by analyzing the angle-dependent optical spectrum of a GX under the photo-excitation of VVB, which is associated with the $\vect{Q}$-dependence of $\Gamma_{G, \vect{Q}}^{\sigma, \ell, \sigma' \ell'}$, \cite{schneider2020direct} thereby inferring the optically transferred TAM in the excited GX state.
The effect made by the last cross-term of Eq. (\ref{TGX}) is especially pronounced as the functional product of $F_{|\ell|}(Q)F_{|\ell'|}(Q)$ and the factor $\sin\beta$ is significantly valued. The two form factors are maximized as $\ell=- \ell'$ and $\beta=\pi/2$. For a VVB with $\ell=-\ell'$ and $\sigma =- \sigma'$, the winding number for the cross term,  $n=\vert\Delta J \vert= \vert\ell'+\sigma'-\ell-\sigma \vert= 2\vert\ell + \sigma\vert$.  
 Figure \ref{Fig5}(a) shows the squared magnitudes of the transition rates, $\Gamma_{B,\vect{Q}}^{+1, 1, -1, -1} (\alpha , \beta) $, of the BX doublet under the excitation of the VVBs formed by the superposition of the TLs, $|1, 1\rangle $ and $|-1, -1 \rangle$, with the different geometric angles,  $(\alpha,\beta)=(0,0)$, $(0, \pi)$, $(0, \pi/2)$ and $(\pi, \pi/2)$. The four selected VVBs are indicated by the north pole, south pole, and the two positions at the equator of the high-order Poincar\'e sphere.
Under the excitation of the same VVBs, Figure \ref{Fig5}(b) shows the squared magnitudes of the $\vect{Q}$-dependent transition rates, $\Gamma_{G,\vect{Q}}^{+1, 1, -1, -1}  $, for the GX states.

As expected from the preceding analysis, the donut-like distribution of the  $\Gamma_{B,\vect{Q}}^{+1, 1, -1, -1} (\alpha, \beta)$ over the $\vect{Q}$-space for the BX doublet under the excitation of the superposition TLs remains invariant against the varied $\alpha$ and $\beta$  (see Supporting Information).   
By contrast, the distribution of the $\Gamma_{G,\vect{Q}}^{+1, 1, -1, -1} (\alpha, \beta)$ over the $\vect{Q}$-plane for the GX states varies with changing the geometric angles, $\alpha$ and $\beta$. 
In particular, at the equator ($\beta=\pi/2$) where the VVB is the maximal superposition of TLs, the $\phi_{\vect{Q}}$-varying patterns of $\Gamma_{G,\vect{Q}}^{+1, 1, -1, -1} (\alpha, \pi/2)$, exhibits the anisotropic patterns with the four-fold ($n=4$) rotation symmetry that directly reflect $\vert \Delta J \vert = 4$ carried by the incident VBB. Generalizing the analysis for the TLs carrying arbitrary OAM, one can show that the transferred OAM to a GX can be inferred from the $n-$fold pattern of $\Gamma_{G,\vect{Q}}^{+1, \ell, -1, -\ell} (0, \pi/2)$ according to the formulation,
\begin{align}
|\ell|= (n-2)/2 \, .
\label{ell}
\end{align}
 The calculated $\vect{Q}$-dependent patterns of   $\Gamma_{G,\vect{Q}}^{+1, \ell, -1, -\ell} (0 , \pi/2)$ for the GX states excited by the VVBs in the higher order modes with $\ell=2,3,4$ are presented in Fig. S4 of Supporting Information, confirming the formalism for extracting the transferred angular momentum from the $n$-fold rotational symmetry of the $\vect{Q}$-dependent pattern of the magnitudes of the transition rates of the GXs by VVBs. 
 Note that, in this manner, the optical information of OAM is imprinted in the $n$-fold petal-like pattern of the squared magnitude of the transition rates of GXs over the $\vect{Q}$-plane, which is measurable by the angle-resolved optical spectroscopy, with no need of good coherence in materials.

\begin{figure*}
	    \centering
        \includegraphics[width=\textwidth]{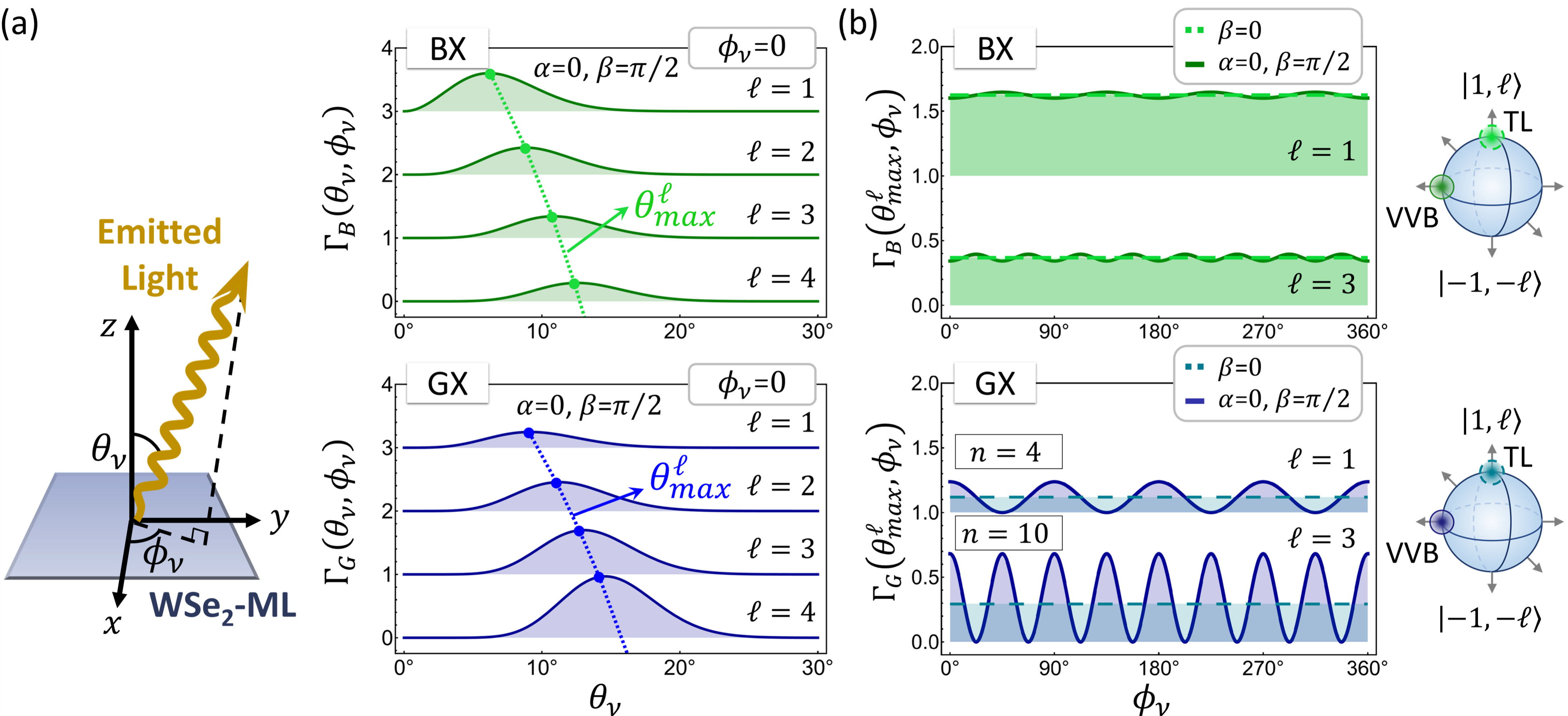}
	    \caption{The angle-dependent optical transition rate, $\Gamma_S(\theta_\nu,\phi_{\nu})$, of the BX doublet and GX states excited by the $\ell$-encoded VVB and single TL. (a) The $\theta_\nu$-dependence of the transition rate, $\Gamma _{S} (\theta _{\nu} , \phi _{\nu} = 0)$, for the BX (upper) and GX states (lower) photo-excited by the $\ell$-encoded VVB state $\vert \sigma , \ell , -\sigma , -\ell ; \alpha , \beta \rangle$ along the $x$-$z$ plane ($\phi _{\nu} = 0$). Note that the polar angle, $\theta_{max}^{\ell}$, for the maximum transition rate increases with the increasing value of $\ell$ in the VVB. The schematic on the left-hand side illustrates the polar, $\theta_\nu$, and azimuthal angle, $\phi_\nu$, of the light emitted from a WSe$_{2}$-ML. (b) The $\phi_{\nu}$-dependence of the transition rate, $\Gamma _{S} (\theta _{\nu} = \theta _{max} ^{\ell} , \phi_{\nu})$, for the BX (upper) and GX states (lower) exited by the $\ell$-encoded single TL (i.e. the state with $\beta=0$ on the Poincaré sphere) and VVB with $(\alpha=0,\beta=\pi/2)$ at $\theta _{\nu} = \theta _{max} ^{\ell}$. Note that the $\phi _{\nu}$-dependence of the transition rate of the GX states mainly follow the $n$-fold patterns appeared in the $\phi _{\vect{Q}}$-dependence of the transition rates in Fig. \ref{Fig5}(b).}
	    \label{Fig6}
\end{figure*}

\subsection*{Read-Out of Imprinted OAM: Angle-Resolved Optical Spectroscopy}
The momentum-dependent transition rates presented in Fig. \ref{Fig5} can be related to the angle-dependent optical pattern. \cite{peng2022tailoring} Accordingly, the feasibility of using the angle-resolved optical spectroscopy \cite{schneider2020direct,peng2022tailoring,kunin2023momentum, PhysRevLett.115.176801} to infer the imprinted optical information in the GXs photo-generated by a VVB is suggested.

Following the theory of Section III.B in Ref. \citen{peng2022tailoring}, the angle-dependent transition rate of the light emission from an BX and GX state is derived as 
\begin{align}
    \Gamma _{B} (\theta _{\nu} , \phi _{\nu}) \propto
    &\left\vert \tilde{M} _{B + , \vect{Q}} ^{\sigma , \ell} \right\vert ^{2} \left\vert D _{B + , \vect{Q}} ^{X} \right\vert ^{2} \left[ 1 -\left( \frac{Q}{Q _{c}} \right) ^{2} \right] + \nonumber \\
    &\left\vert \tilde{M} _{B - , \vect{Q}} ^{\sigma , \ell} \right\vert ^{2} \left\vert D _{B - , \vect{Q}} ^{X} \right\vert ^{2} \label{Emission_Rate_B}
\end{align}
and 
\begin{align}
    \Gamma _{G} (\theta _{\nu} , \phi _{\nu}) \propto
    \left\vert \tilde{M} _{G , \vect{Q}} ^{\sigma , \ell} \right\vert ^{2} \left\vert D _{G , \vect{Q}} ^{X} \right\vert ^{2} \left( \frac{Q}{Q _{c} ^{\prime}} \right) ^{2} \, , \label{Emission_Rate_G}
\end{align}
respectively, where $\theta _{\nu}$ and $\phi _{\nu}$ are the polar and azimuthal angles of the emitted plane-wave light, respectively. In the equations, the angles, $\theta _{\nu}$ and $\phi _{\nu}$,  of emitted light are related to the vectorial momentum of exciton, $\vect{Q}= Q (\cos\phi_{\vect{Q}} \hat{\vect{x}} + \sin\phi_{\vect{Q}} \hat{\vect{y}})$, following the conservation laws of momentum and energy.
The momentum conservation law ensures that $\phi _{\nu} = \phi _{\vect{Q}}$, while the energy conservation law leads to $\theta _{\nu} = \sin ^{-1} (Q / Q _{c})$, where $Q _{c} = E _{S, \vect{0}} ^{X} / \hbar c$ is the light-cone radius. Because of the different energies, the wavevectors of light-cone edges for a BX and GX slightly differ, which are given by $Q _{c} = E _{B \pm , \vect{0}} ^{X} / \hbar c$ and $Q _{c} ^{\prime} = E _{G , \vect{0}} ^{X} / \hbar c \approx Q_c$, respectively.

In Fig. \ref{Fig6}, we show the angle-dependent optical transition rates of the BX doublet and GX state excited by the $\ell$-encoded VVBs, $\vert \sigma,\ell, -\sigma, -\ell ; \alpha , \beta \rangle$, with different values of $\ell$. Apparently, one sees that the azimuthal-angle-dependence of the transition rate of the GX state reflect the $n$-fold rotation symmetry of the squared magnitude of the transition rate of the GX state photo-excited by the VVB in the equator state  presented in Fig. \ref{Fig5}(b).
As the transition dipoles of GXs are weakly $\vect{Q}$-dependence, according to Eq. (\ref{Emission_Rate_G}), the azimuthal-angle-dependence of the optical pattern of an GX mainly follow the $\phi_{\vect{Q}}$-dependence of the transition rate. Hence, it is suggested to read out the imprinted optical information in VVB-excited GX states by analyzing the azimuthal-angle-dependences of the optical patterns that are measurable by angle-resolved optical spectroscopy.\cite{schneider2020direct,peng2022tailoring,kunin2023momentum, PhysRevLett.115.176801}

\section*{CONCLUSION}
In summary, we have presented a comprehensive investigation based on first-principles, focusing on the light-matter interaction between structured lights carrying optical angular momenta and tightly bound excitons in 2D materials. We show that the photo-excitation of a specific type of spin-forbidden dark excitons,  i.e. gray exciton, is greatly enhanced by the incident twisted lights that carry orbital angular momentum and  possess the longitudinal field component associated with the interaction between spin and orbital angular momenta. Moreover, we investigate the superposition of two twisted lights with distinct SAM and OAM, resulting in the formation of a vector vortex beam (VVB) that is spatially engineered in both complex amplitude and polarization as well.  Our research demonstrates that a spin-orbit-coupled VVB in a non-separable form surprisingly allows for the imprinting of the carried optical information onto gray excitons in 2D materials, which is robust against the decoherence mechanisms in materials. These studies unveil the indispensable role of gray excitons in twisted-light-based optoelectronics and suggest the utilization of VVB for transferring optical information onto  2D materials.

\section*{METHODS}
\textbf{Density Functional Theory calculations}. The Density Functional Theory (DFT) calculations were conducted using the PBE functional \cite{perdew1996generalized} as implemented in the Quantum Espresso software package. \cite{giannozzi2009quantum} The supercell was constructed with a 30 Å-high vacuum in the aperiodic direction and an in-plane lattice constant of 3.35 Å. A plane-wave cutoff energy of 1632 eV was employed for the expansion of wavefunctions, along with norm-conserving pseudo-potentials. The electronic structure calculations utilized a Monkhorst-Pack grid of $9 \times 9 \times 1$ points to sample the first Brillouin zone. Convergence of the electronic self-consistent loop was achieved with a break condition set to $1.36 \times 10^{-9}$ eV.

\section*{ASSOCIATED CONTENT}
\subsection*{Supporting Information} 
Technical details on establishing the Bethe-Salpeter equation based on first-principles; guidelines for evaluating the transition dipole of an exciton; formalisms for Laguerre-Gaussian beams in both real and momentum space under the Coulomb gauge; demonstration of the encoding of various angular momentum in the transition rates of gray excitons.

\section*{ACKNOWLEDGMENTS}
This study is supported by National Science and Technology Council of Taiwan, under contracts, MOST 109-2112 -M-009 -018 -MY3.

\bibliographystyle{achemso.bst}
\bibliography{references.bib}

\end{document}